**Origin of superconducting carriers in "non-doped" *T'*- (La, *RE*)$_2$CuO$_4$ (*RE* = Sm, Eu, Gd, Tb, Lu, and Y) prepared by molecular beam epitaxy**

M. Noda [a, b], A. Tsukada* [b], H. Yamamoto [b], M. Naito [c]

[a] Tokyo University of Science, 2641 Yamazaki, Noda, Chiba 278-8510, Japan

[b] NTT Basic Research Laboratories, NTT Corporation, 3-1 Wakamiya Morinosato, Atsugi-shi, Kanagawa 243-0198, Japan

[c] Department of Applied Physics, Tokyo University of Agriculture and Technology, 2-24-16 Naka-cho, Koganei, Tokyo 184-8588, Japan

**Abstract**

We have performed a systematic investigation of the variations of the lattice constants with substituent rare-earth element concentration $x$ in the nominally undoped superconductors *T'*-La$^{3+}_{2-x}$*RE*$^{3+}_x$CuO$_4$ (*RE* = Sm, Eu, Gd, Tb, Lu, and Y), which we have recently discovered using MBE.  The results show both the in-plane and out-of-plane lattice constants ($a_0$ and $c_0$) linearly decrease with $x$, whose extrapolation to $x$ = 2 agrees well with the reported $a_0$ and $c_0$ values for each *T'*-*RE*$_2$CuO$_4$.  This behavior is what one would expect simply from the ionic size difference between La$^{3+}$ and *RE*$^{3+}$.  The absence of the Cu-O bond stretching due to electron-doping, which is commonly observed in electron-doped *T'* and infinite-layer superconductors, implies that electron doping *via* oxygen deficiencies is , at least, not a main source of charge carriers.

*Corresponding author.

Dr. Akio Tsukada

Postal address: Thin-Film Materials Research Group, NTT Basic Research Labs., 3-1 Morinosato-Wakamiya, Atsugi-shi, Kanagawa 243-0198, Japan

Phone: +81-46-240-3349

Fax: +81-46-240-4717

E-mail address: tsukada@will.brl.ntt.co.jp

1. **Introduction**

It has commonly been believed that the high-$T_c$ superconductivity develops in a Mott-insulating mother compound by doping either hole or electron carriers. However, our recent observation of fairly high-$T_c$ superconductivity (> 20 K) in nominally undoped $T'$-La$^{3+}_{2-x}RE^{3+}_{x}$CuO$_4$ ($RE$ = rare earth elements) [1-3] appears to contradict this general belief. The purpose of this paper is to examine on the basis of a systematic investigation of the variation of the lattice constants with $x$ whether these materials are really undoped or not. One may think there is a possibility that oxygen deficiencies outside the CuO$_2$ planes serve as a source of effective electron carriers. The most straightforward approach to resolving this issue is to determine oxygen content very precisely along with its site-specific occupancy. However, with present technologies, this is impossible, especially for thin film specimens. An alternative approach is to investigate the variation of lattice parameters, especially the in-plane lattice constant $a_0$, with substituent $RE$ concentration $x$. It is well known that electron-doping stretches and hole-doping shrinks the Cu-O bond length due to accumulation or depletion of electrons in the Cu-O $dp\sigma_{x^2-y^2}$ anti-bonding bands [4-6]. Hence, $a_0$ could be a useful measure of doping. To examine this, we have performed a systematic investigation of the variations of the lattice constants in $T'$-La$^{3+}_{2-x}RE^{3+}_{x}$CuO$_4$ ($RE$ = Sm, Eu, Gd, Tb, Lu, and Y) thin films with changing $x$. The results show both the $a_0$ and $c_0$ values vary in a manner that one would expect by taking into account only the geometrical effect due to the ionic size difference between La$^{3+}$ and $RE^{3+}$. The absence of the Cu-O bond stretching implies that electron-doping *via* oxygen deficiencies is, at least, not a main source of charge carriers.

## 2. Experimental

We grew $La_{2-x}RE_xCuO_4$ ($RE$ = Sm, Eu, Gd, Tb, Lu, and Y) thin films in a customer-designed molecular beam epitaxy (MBE) chamber from metal sources using multiple electron-gun evaporators with accurate stoichiometry control of the atomic beam fluxes [7]. The substrate temperature was typically ~ 650°C, and during growth 1 - 5 sccm of ozone gas (10% $O_3$ concentration) was supplied to the substrate for oxidation. The growth rate was ~1.5 Å/s, and the film thickness was typically ~ 900 Å. After the growth, the films were held at about 630°C in vacuum ($p_{O_2} < 10^{-8}$ Torr) for 10 minutes to remove interstitial apical oxygen. For the $T'$-phase stabilization, we mainly used $YAlO_3$ (100) substrates and sometimes $KTaO_3$ (100), $NdCaAlO_4$ (001), or $SrTiO_3$ (100) substrates [8]. The crystal structures and the lattice constants of the grown films were determined using a 4-circle X-ray diffractometer.

## 3. Results and discussion

First, we show the results of X-ray diffraction (XRD) measurements. Figure 1 shows a typical $\theta$-$2\theta$ scanned XRD pattern of the $T'$-$La_{1.85}Y_{0.15}CuO_4$ film. All the peaks can be indexed to $002n$ reflections, indicating the film is $c$-axis oriented. The lattice constant is calculated to be $c_0$ = 12.43 Å. The film is also in-plane oriented based on 4-circle XRD and reflection high-energy electron diffraction (RHEED) experiments. For $RE$ = Sm, Eu, Gd, and Tb, the single $T'$ phase was obtained essentially for the whole range of substitution ($0 < x \leq 2$), whereas for $RE$ = Y and Lu, the $T'$ phase were obtained for $x \leq 1.0$ and $x \leq 0.3$ under the present growth conditions [2]. Figure 2 shows the variations of the lattice constants with $x$ in the $T'$-$La_{2-x}RE_xCuO_4$ ($RE$ = Sm, Eu, Gd, Tb, Lu, and Y) films grown on $NdCaAlO_4$ ($x$ = 0),

YAlO$_3$ (0 < $x$ ≤ 1), and SrTiO$_3$ substrates (1 < $x$ ≤ 2). Both the $a_0$ and $c_0$ decrease monotonically with increasing $x$ in every RE substitution. The extrapolation of $a_0(x)$ and $c_0(x)$ to $x = 0$ yields the $a_0$ and $c_0$ of T'-La$_2$CuO$_4$ stabilized by low-temperature (~550°C) growth [8], and the extrapolation to $x = 2$ yields the $a_0$ and $c_0$ values of T'-RE$_2$CuO$_4$ [9, 10]. This behavior is what one would expect simply from the ionic size difference between La$^{3+}$ and RE$^{3+}$. In this figure, the superconducting films are represented by filled symbols; specifically, 0.15 ≤ $x$ ≤ 0.5 for Sm, 0.15 ≤ $x$ ≤ 0.3 for Eu, 0.2 ≤ $x$ ≤ 0.5 for Gd, 0.1 ≤ $x$ ≤ 0.3 for Tb, $x$ = 0.07 for Lu, and 0.1 ≤ $x$ ≤ 0.2 for Y [2], suggesting a correlation between $a_0$ and the appearance of superconductivity. It should also be mentioned that although the partial oxygen pressure during "reduction" after the growth significantly affects the superconducting properties of the films, it affects the lattice constants very little, which we discuss elsewhere [11]. Figure 3 shows the variations of $a_0$ and $c_0$ in T'-La$^{3+}_{2-x}$Gd$^{3+}_x$CuO$_4$ in comparison with that of "electron-doped" T'-La$^{3+}_{2-x}$Ce$^{4+}_x$CuO$_4$ [2, 12, 13] on an enlarged scale. The $c_0$ decreases linearly with $x$ in both materials, simply reflecting the difference in ionic radius of Gd$^{3+}$ ($r_i$ = 1.053 Å) or Ce$^{4+}$ ($r_i$ = 0.970 Å) versus La$^{3+}$ ($r_i$ = 1.160 Å) [14]. In contrast, the $a_0(x)$ curves are different: $a_0$ of T'-La$_{2-x}$Gd$^{3+}_x$CuO$_4$ linearly decreases with $x$, whereas $a_0$ of T'-La$_{2-x}$Ce$^{4+}_x$CuO$_4$ shows complicated behavior; namely, it initially decreases and then starts to increase slightly above $x > 0.08$ in spite of the substantially small Ce$^{4+}$ as compared to La$^{3+}$. The complicated $a_0(x)$ behavior in La$_{2-x}$Ce$^{4+}_x$CuO$_4$ arises from a delicate balance between the Cu-O bond stretching by electron doping and the geometrical effect by substitution of Ce$^{4+}$ for La$^{3+}$.

Next, we discuss the origin of superconducting carriers in T'-La$_{2-x}$RE$_x$CuO$_4$. Assuming that RE$^{3+}$ ions do not have a deviating valency, which seems to be a reasonably safe assumption [1], one can think of three possible scenarios: the materials are (i) hole-doped, achieved by excess oxygen like $T$-La$_2$CuO$_{4+\delta}$; (ii) electron-doped, achieved by oxygen deficiencies at the O(2) site [regular oxygen site in the (La, $RE$)$_2$O$_2$ plane]; or (iii) truly undoped [15]. It is safe to rule out the first possibility because it is well known that, in $T'$-structure superconductors, the excess oxygen occupies apical sites and *never* leads to metallicity nor superconductivity [8, 16, 17]. With regard to the second possibility, the $a_0(x)$ curves of $T'$-La$_{2-x}RE_x$CuO$_4$ as shown in Fig. 2 and Fig. 3 argue against substantial electron doping, although the present experiments cannot completely rule out a tiny amount of oxygen deficiencies at O(2), which might be introduced in the reduction process for removing the residual impurity oxygen at apical sites. A comparison of the bond energies between Cu-O (~270 kJ/mole) and $RE$-O (~420 – 800 kJ/mole) molecules [18] also suggests it *unlikely* that only O(2) sites are oxygen deficit with O(1) sites intact. Furthermore, our *in-situ* photoemission spectroscopy and transport experiments also disagree with the second scenario [2, 11]. Briefly, the valence band spectra showed a rigid-band shift of ~ 0.07 eV between undoped La$_{1.8}$Eu$_{0.2}$CuO$_4$ and electron-doped La$_{1.9}$Ce$_{0.1}$CuO$_4$, which agrees with the chemical potential shift observed in La$_{2-x}$Ce$_x$CuO$_4$ (~ 0.7 eV per Ce) [2]. In the transport measurements, La$_{1.85}$Y$_{0.15}$CuO$_{4+y}$ films reduced in different conditions, eventually with different $y$, showed no indication of an expected Mott-Hubbard transition to a highly insulating state around $y \sim 0$ [11]. Therefore, at present, we believe that the third scenario is the most plausible. In square-planar cuprates, apical oxygen is known to be a strong pair breaker and a strong scatterer. A slight amount of remnant apical oxygen atoms severely distort the generic phase diagram [19, 20], and also lead to the *apparent* semiconducting state in $T'$ mother compounds such as $Pr_2CuO_4$ and $Nd_2CuO_4$. In $T'$-$La_{2-x}RE_xCuO_4$, a more thorough removal of apical oxygen atoms is achieved due to the longer $a_0$ than in $Pr_2CuO_4$ and $Nd_2CuO_4$, which unveils the generic electronic state of the $T'$ compounds for the first time [1, 3, 13].

## 4. Summary

We have systematically performed a series of lattice constant measurements of $T'$-$La^{3+}_{2-x}RE^{3+}_xCuO_4$ ($RE$ = Sm, Eu, Gd, Tb, Lu, and Y) to study the origin of superconducting carriers in these nominally undoped superconductors. The results indicate that both $a_0$ and $c_0$ monotonically decrease with $x$, whose extrapolation to $x = 2$ agrees well with the reported $a_0$ and $c_0$ values of each $T'$-$RE_2CuO_4$, simply reflecting the ionic size difference between $La^{3+}$ and $RE^{3+}$. The absence of the Cu-O bond stretching implies that electron doping *via* oxygen deficiencies is, at least, not a main source of charge carriers. These new superconductors are most plausibly "*band superconductors*".

**Acknowledgements**

The authors are grateful indeed to Prof. L. Alff of Vienna University and Mr. Y. Krockenberger of Max-Plank-Institute for Solid State Research for fruitful discussions. They also thank Dr. Y. Taniyasu for his help in XRD measurements, Dr. T. Yamada, Dr. H. Sato, Dr. H. Shibata, Dr. S. Karimoto, Dr. K. Ueda, Dr. J. Kurian, and Dr. A. Matsuda for helpful discussions, and Dr. T. Makimoto for his support and encouragement.

[1] Even for Tb, where mixed valency ($Tb^{3+}/Tb^{4+}$) occurs in other systems, the $Tb^{3+}$ state is clearly established by the results of X-ray photoelectron spectroscopy [3].

**Figure captions**

Fig. 1. XRD pattern ($\theta$-$2\theta$ scanned) of MBE-grown $T'$-$La_{1.85}Y_{0.15}CuO_4$ thin film. The unindexed peaks are due to the substrate.

Fig. 2. Variation of lattice constants with substituent $RE$ concentration $x$ for $T'$-$La_{2-x}RE_xCuO_4$ ($RE$ = Sm, Eu, Gd, Tb, Lu, and Y) thin films. Squares and circles denote $a_0$ and $c_0$, respectively. Closed symbols indicate the films are superconducting. The reported $a_0$ and $c_0$ values of bulk $T'$-$RE_2CuO_4$ specimens are also plotted (white crosses and crosses, respectively) [9, 10].

Fig. 3. The $a_0$ and $c_0$ of $T'$-$La^{3+}_{2-x}Gd^{3+}_xCuO_4$ (squares) and $T'$-$La^{3+}_{2-x}Ce^{4+}_xCuO_4$ (circles) as a function of Gd or Ce concentration $x$. Lines are guides for the eye.

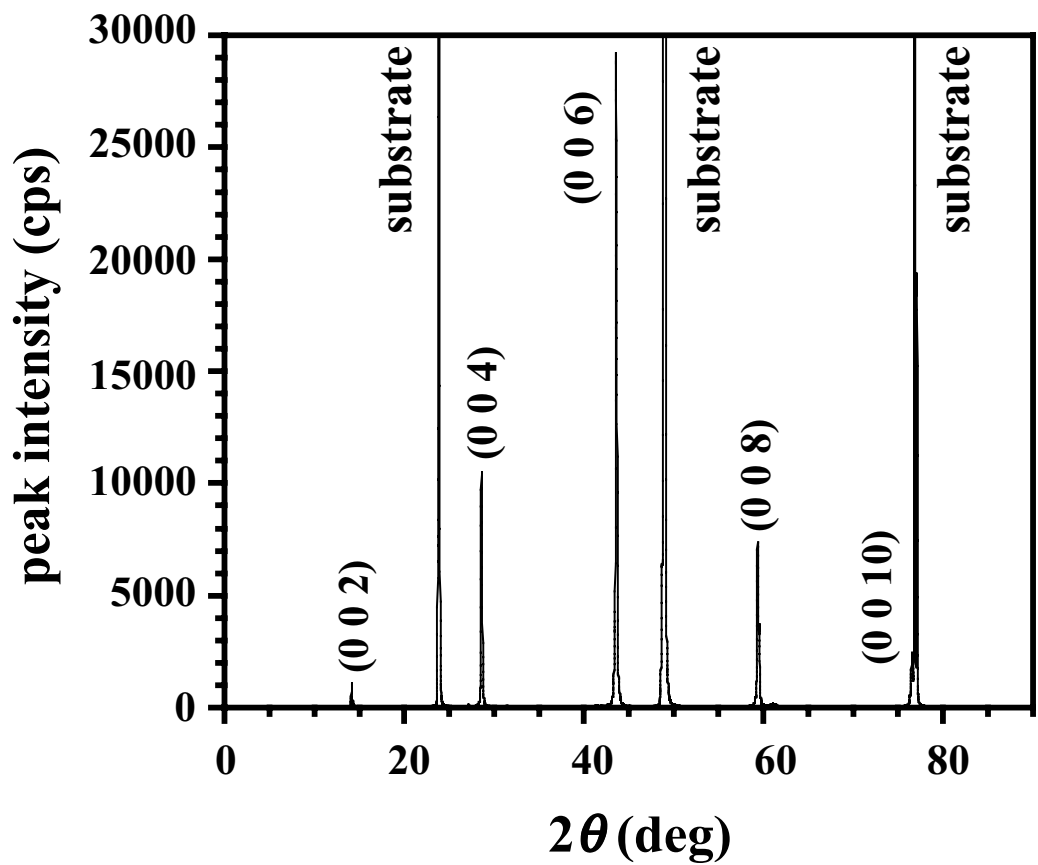

**Figure 1 M.Noda et al.**

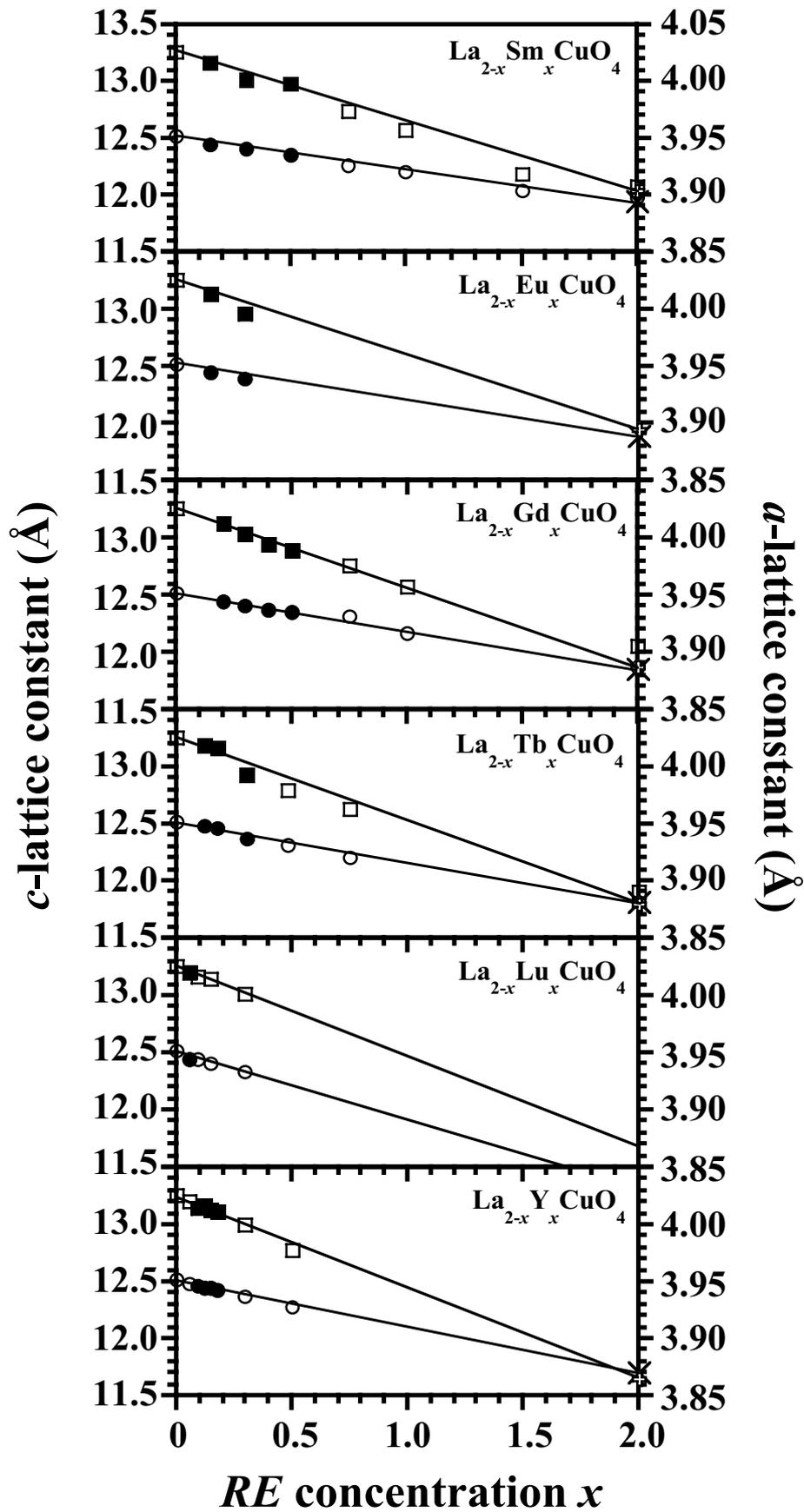

Figure 2  M.Noda et al.

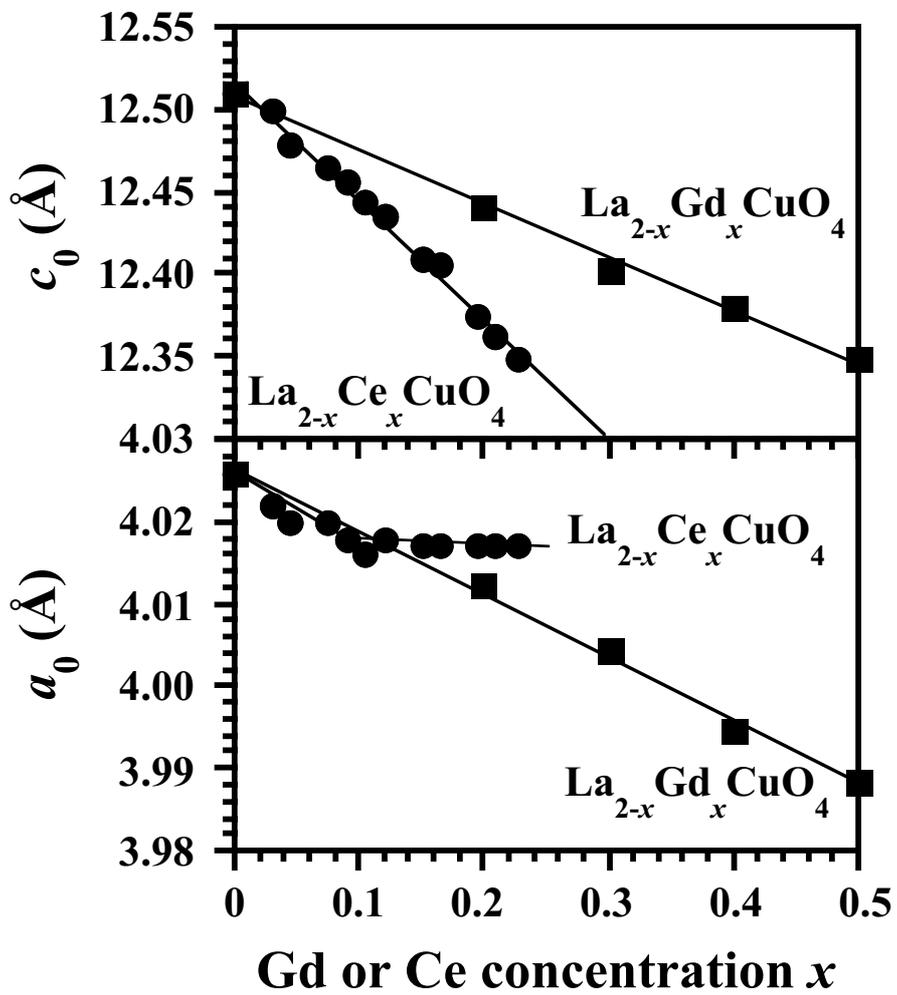

**Figure 3  M. Noda et al.**